\documentclass[11pt]{article}

\usepackage{moriond}
\usepackage{graphicx}

\def \snu  { \mathrm{\tilde{\nu}}}
\def\21{SU(2) $\otimes$ U(1) }

\newcommand{\TeV}{\, {\rm TeV}}

\begin{document}
\vspace*{4cm}
\title{Reconciling dark matter and neutrino masses in mSUGRA}

\author{ Chiara Arina }

\address{Service de Physique Th\'eorique, Universit\'e Libre de Bruxelles,\\
  CP225, Bld du Triomphe, 1050 Brussels, Belgium}

\maketitle\abstracts{We study the minimal SUGRA phenomenology in the case of an alternative seesaw mechanism for generating neutrino masses. Changes in the neutrino sector lead to a modification of the supersymmetric particle spectrum and the sneutrino naturally arises as the lightest supersymmetric particle. The obtained sneutrino has a relic density within the WMAP range and is compatible with present nuclear recoil bounds.}

\section{Introduction}

The experimental evidences for neutrino masses and oscillations on one hand and the need of non baryonic Dark Matter (DM) from cosmological studies on the other hand are indications for physics beyond the standard model. We wish to reconsider in a consistent way sneutrino as a cold relic from the early Universe and study its phenomenology relevant both for Cosmology and for
relic-particle detection in connection with the generation of neutrino masses.

The sneutrino as dark matter candidate has been widely studied, in different supersymmetric models. In the Minimal Supersymmetric Standard Model (MSSM), sneutrinos are only left-handed, being the neutrino superpartners. In terms of DM candidates they are marginally compatible with direct detection bounds, provided they compose a subdominant component of dark matter. Such incompatibility is mainly due to the coupling between the $Z$ boson and the sneutrino, leading to a relic density below the WMAP range and to high scattering cross-sections on the detector nuclei. Possible ways to weak this coupling are the mixing of the left-handed sneutrino through a sterile right-handed field and/or the introduction of lepton-number violating terms. Furthermore neutrinos in the MSSM are massless.

A more detailed and extended analysis of the sneutrino phenomenology in connection with the neutrino physics in non minimal effective supersymmetric models is presented in~\cite{Arina:2007tm} and references therein, whereas the inverse seesaw model and its outcomes on the sneutrino sector is described in~\cite{Arina:2008bb}. 

\section{Mixed sneutrinos as cold dark matter (CDM) candidates}
The minimal supersymmetric model requires an extension in order to provide mass to neutrinos. Models with lepton-number violation terms can allow for Majorana neutrino masses. The most direct way to include a Majorana mass term is to introduce a non-renormalizable gauge-invariant dimension-5 operator of the type $\mathcal{L}=g_{IJ}/M_{\Lambda}(\epsilon_{ij} L_i^I H_j)\epsilon_{kl}L_k^J H_l) +\rm{h.c.}$, from~\cite{Hall:1997ah,Arkani-Hamed:2000bq,Hirsch:1997vz}, where $L_i$ are the left-handed neutrino superfield. In this case, a Majorana mass term for the neutrino is generated when the neutral component of the Higgs field, $H_j$ acquires a vacuum expectation value and the neutrino mass which arises is of the order of $m_{M} \sim gv^2/M_{\Lambda}$. This can be made compatible with neutrino mass bounds for $M_{\Lambda}$ close to the GUT scale. Lepton-number violating terms are now allowed in the sneutrino potential as well, but they do not lead to a significant modification of the sneutrino phenomenology respect to the MSSM when the neutrino mass bounds are properly included. 

On the contrary superpartner phenomenology is greatly altered by the presence of weak scale right-handed sneutrinos and additional singlet fields and may provide naturally to a mixed sneutrino as DM candidate. We want to show that models with right-handed neutrino singlet fields are perfectly viable, especially if embedded with various type of seesaw mechanisms. For a standard seesaw mechanism, discussed in Sec.~\ref{subsec:ssm} for explaining the neutrino physics, the analysis is done at the electroweak scale without imposing unification of scalar soft masses: the sneutrino can be the lightest supersymmetric particle (LSP) and a good DM candidate. In Sec.~\ref{subsec:invssm} we then describe the inverse seesaw model for generating neutrino masses, which nicely accommodate the framework of a minimal supergravity theory (mSUGRA) with a sneutrino dark matter.

\subsection{Seesaw model: implications in the sneutrino sector}\label{subsec:ssm}

A supersymmetric model which can accommodate both Dirac and Majorana mass--terms for neutrinos and explain the 
observed neutrino mass pattern, and which relies on a renormalizable lagrangian, may be built by adding to the minimal MSSM right--handed fields $\tilde N_{i}$ and allowing for lepton-number violating terms. The most general form of the superpotential and of the soft supersymmetry--breaking potential which accomplishe these conditions are~\cite{Arkani-Hamed:2000bq,Grossman:1997is}:
\begin{eqnarray}
\mathcal{W} & = & {\cal W_{\rm MSSM}} +  \varepsilon_{ab}\,
Y_{\nu}^{ij}\widehat L_i^a\widehat N_j\widehat H_2^b+ \frac{1}{2} M^{ij} \hat N_i \hat N_j\\
V_{\rm soft} & = & V_{\rm soft}^{\rm MSSM} + (M_{N}^{2})_{ij} \, \tilde N^{\ast}_{i} \tilde N_{j} -
 [(m^2_B)_{ij}\tilde{N}_i\tilde{N}_j+\epsilon_{ab} A_{\nu_{ij}} H^{a}_{2} \tilde L_{i}^{b} \tilde N_{j}  + \mbox{h.c.}]\nonumber
\end{eqnarray}
where ${\cal  W_{\rm MSSM}}$  is the usual MSSM superpotential, $V_{\rm soft}^{\rm MSSM}$ is the MSSM SUSY--breaking scalar potential and $M^{ij}$, $Y_{\nu}^{ij}$,  $(M_{N}^{2})_{ij}$, $(m_{B}^{2})_{ij}$, and $A_{\nu_{ij}}$ are matrices which we choose diagonal in flavour space. For the lepton-number violating parameters we therefore assume: $M^{ij}=M\;\delta^{ij}$. The Dirac mass of the neutrinos is obtained as: $m_D^{i} = v_{2}Y_{\nu}^{ii}$, with $Y_{\nu}$ the neutrino Yukawa coupling. $M$ represents a Majorana mass--term for neutrinos and the neutrino mass is defined as usual through the seesaw mechanism as $m^{\rm eff}_{\nu}\simeq m_{D}^{i2}/M$. Sneutrinos now are a superpositions of two complex fields: the left--handed field $\tilde{\nu_{L}}$ and the right--handed field $\tilde{N}$. Since we introduced Majorana terms, it is convenient to work in a basis of CP eigenstates, therefore the mass matrix in the vector basis $\Phi^{\dagger}= \left(\snu_+\ \tilde{N}_+\ \snu_-\ \tilde{N}_-\right)$ is: 
{\small
\begin{eqnarray}
\mathcal{M}^2_{Maj}  = 
\left (
\begin{array}{cccc}
m^2_L+D+m^2_D & F^2 + m_D M & 0  & 0\\
F^2+m_D M & m^2_N+M^2+m^2_D+m^2_B & 0 & 0\\ 
0 & 0 & m^2_L+D+m^2_D & F^2-m_D M \\
0 & 0 & F^2- m_D M & m^2_N+M^2+m^2_D-m^2_B
\end{array}
\right )
\label{eq:majmass}
\end{eqnarray}}
with $D^2 = 0.5\, m_{Z}^{2} \cos(2\beta)$ and $F^2= v A_\nu \sin\beta - \mu m_D\rm cotg\beta$. Sneutrino mass eigenstates are obtained by diagonalizing Eq. \ref{eq:majmass}. We define them as follows:
\begin{eqnarray}
\snu_i= Z_{i1}\snu_{+}+Z_{i2}\tilde{N}_+ +Z_{i3}\snu_{-}+Z_{i4}\tilde{N}_{-} \qquad i=1,2,3,4
\end{eqnarray}
The lightest state, which is our dark matter candidate, may now exhibit a mixing with the right--handed
field $\tilde N$ and the non--diagonal nature of the $Z$--coupling with respect of the CP eigenstates, therefore his interaction with the $Z$ boson is highly reduced. 

\begin{figure}[t!]
\vspace{-1cm}
\begin{minipage}[t]{0.5\textwidth}
\centering
\includegraphics[width=\columnwidth]{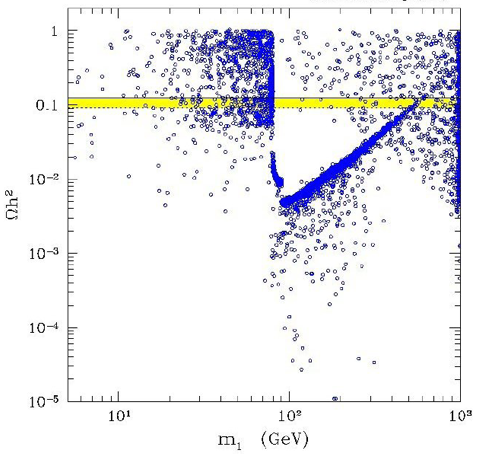}
\caption{Sneutrino relic abundance $\Omega h^{2}$ as a function of the sneutrino mass $m_{1}$ for the case of a Majorana-mass parameter $M = 1$ TeV and for a full scan of the supersymmetric
parameter space. All the models shown in the
plot are acceptable from the point of view of all experimental constraints. The yellow band delimit the three years WMAP interval for CDM.}
\label{fig:om_majA}
\end{minipage}
\hspace{0.5 cm}
\begin{minipage}[t]{0.5\textwidth}
\centering
\includegraphics[width=\columnwidth]{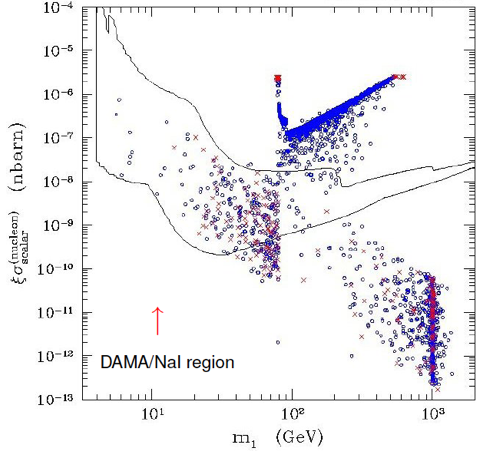}
\caption{ Sneutrino scattering on nucleon $\xi\sigma_{nucleon}^{(scalar)}$ as a function of the sneutrino mass $m_1$. [Red] crosses denote sneutrino configurations with the relic density within the WMAP range, [blue] open points refer to cosmologically subdominant sneutrinos. The black line delimits the experimental DAMA/NaI annual modulation region.}
\label{fig:sig_majA}
\end{minipage}
\end{figure}

The sneutrino relic abundance is computed considering all possible annihilation channels and taking into account coannihilation with the charged sleptons and with the heavier sneutrino eigenstate $\snu_2$, when occurring. We performed a scan over the usual MSSM parameters and over the free parameters which appear in Eq.~\ref{eq:majmass} for the sneutrino sector. More details on the value of the supersymmetric parameters can be found in~\cite{Arina:2007tm}. We notice that the phenomenology of the sneutrino becomes cosmologically relevant if the Majorana mass is at the electroweak scale, namely $M \simeq 1 \TeV$ , instead of assuming the typical values of a seesaw. With a Majorana mass of the order of $10^{14}$ GeV, the right-handed sneutrino sector is completely decoupled from the low energy sector, leading to the same phenomenology as in the MSSM. For a Majorana mass of $1$ TeV the sneutrino is compatible with the WMAP three years bound~\cite{wmap} in a wide range of masses, from few GeV up to the TeV, as it is shown in Fig.~\ref{fig:om_majA}.

Regarding the relic particle detection, we focus on signals for direct searches. The direct detection experiments are sensitive to nuclei recoil caused by a WIMP that scatter off of a nucleus in the detector. Sneutrinos are characterized by spin independent interactions, which receive contributions from the $Z$ and Higgs exchange on the t-channel, $\sigma_{\mathcal{N}}=\sigma_{\mathcal{N}}^{Z}+\sigma_{\mathcal{N}}^{h,H}$, detailed in~\cite{Goodman:1984dc,Arina:2007tm}. We consider the scattering cross-section on nucleon, $\xi\sigma^{(scalar)}_{nucleon}$, with $\xi= \min(1,\Omega_{\snu} h^2/\Omega_{CDM}h^2)$ defined as the fractional amount of local non-baryonic DM density.

In the CP basis previously defined, the $Z$--coupling is no longer diagonal, therefore the elastic scattering through t-channel $Z$ exchange, which is dominant channel, becomes an inelastic reaction $\snu_1+N\rightarrow \snu_2+N$. The mixed sneutrino is a nice realization of inelastic dark matter, which was introduced by~\cite{Smith:2001hy}. Since the heavier state must be produced, the direct detection rate is suppressed by a factor $\mathcal{S}$ for kinematical reasons. $\cal{S}$ depends on the sneutrino mass splitting, $\Delta m = m_2 - m_1$, on the recoil energy, on the type of nucleus and on the energy sensitivity of the detector. We thus redefine the scattering cross section to be:
\begin{equation}
\left [\xi \sigma^{\rm (scalar)}_{\rm nucleon} \right ]_{\rm eff} = 
{\cal S} (\xi \sigma^{\rm (scalar)}_{\rm nucleon})^{Z} +  (\xi \sigma^{\rm (scalar)}_{\rm nucleon})^{h,H} 
\label{eq:xisigmareduced}
\end{equation}
The lepton-number violating terms in the lagrangian may induce radiative contributions to the neutrino masses \cite{Hirsch:1997vz,Grossman:1997is}. At 1--loop, these corrections arise from self-energy diagrams involving the sneutrino and neutralino eigenstates and are basically proportional to $\Delta m$. We impose that the radiative contributions do not exceed the experimental upper
bound on the neutrino mass. Sneutrino dark matter phenomenology is therefore
bounded by neutrino physics in a non trivial way, due to the correlation of the direct detection cross-section and the neutrino mass through the mass splitting $\Delta m$. In Fig.~\ref{fig:sig_majA} the scaled cross-section on nucleon versus the sneutrino mass is shown, compared with the DAMA/NaI~\cite{Bernabei:2003za} annual modulation region. Either sneutrino configurations in the WMAP range (red crosses) either cosmologically subdominant sneutrinos (blue circles) are compatible with the experimental results.

\subsection{Inverse seesaw model, mSUGRA unification and sneutrino LSP}\label{subsec:invssm}

In a minimal supergravity scheme where the smallness of neutrino masses is accounted for within the inverse seesaw mechanism, the lightest supersymmetric particle is likely to be represented by the sneutrino, instead of the lightest neutralino. This opens a new window for the mSUGRA scenario. Here we consider the
implications of the model for the dark matter issue. Let us add to the MSSM three
sequential pairs of \21 singlet neutrino superfields $\widehat{N}_i$ and $\widehat{S}_i$, with
the following superpotential terms~\cite{mohapatra:1986bd,Deppisch:2004fa}:
\begin{eqnarray} 
 {\cal W} & = &  {\cal W_{\rm MSSM}}  +\varepsilon_{ab}\,
 Y_{\nu}^{ij}\widehat L_i^a\widehat N_j\widehat H_2^b
+ M_{R}^{ij}\widehat N_i\widehat S_j 
+\frac{1}{2}\mu_S^{ij} \widehat S_i \widehat S_j\\
 V_{\rm soft} &  = & V_{\rm soft}^{\rm MSSM} + (M_{N}^{2})_{ij} \, \tilde N^{\ast}_{i} \tilde N_{j} + (M_{S}^{2})_{ij} \, \tilde S^{\ast}_{i} \tilde S_{j}
+ [\varepsilon_{ab}\,
 A_{h_{\nu}}^{ij}  \tilde{L}_i^a  \tilde{N}_j  H_2^b
+ B_{M_{R}}^{ij}  \tilde{N}_i \tilde{S}_j 
+\frac{1}{2} B_{\mu_S}^{ij}  \tilde{S}_i  \tilde{S}_j + h.c.]\nonumber
\label{eq:Wsuppot} 
\end{eqnarray}
where again all the matrices are chosen diagonal in flavor space. In the limit $\mu_S^{ij} \to 0$ there are exactly conserved lepton
numbers assigned as $(1,-1,1)$ for $\nu$, $N$ and $S$, respectively. Small neutrino masses are generated through the inverse seesaw mechanism~\cite{mohapatra:1986bd,Deppisch:2004fa,Nunokawa:2007qh}: the effective neutrino mass matrix $m^{\rm eff}_{\nu}$ is obtained by the following relation:
\begin{equation}
  \label{eq:1}
  m^{\rm eff}_{\nu}= -v_2^2 Y_{\nu} \left(M_R^T\right)^{-1} {\mu_S}
    M_R^{-1} Y_{\nu}^T = \left(U^T\right)^{-1} m_{\mu}^{\rm diag}\ U^{-1}
\end{equation}
The smallness of the neutrino mass is ascribed to the smallness of the $\mu_S$ parameter, rather than the largeness of
the Majorana--type mass matrix $M_R$, as required in the standard seesaw mechanism~\cite{Nunokawa:2007qh}. In this way light (eV scale or smaller) neutrino masses allow for a sizeable magnitude for the Dirac--type mass and a TeV--scale mass for the
right-handed neutrinos, features which have been shown to produce an interesting sneutrino dark matter phenomenology in the previous section~\ref{subsec:ssm}. 

\begin{figure}[t!]
\vspace{-1cm}
\begin{minipage}[t]{0.5\textwidth}
\centering
\includegraphics[width=\columnwidth]{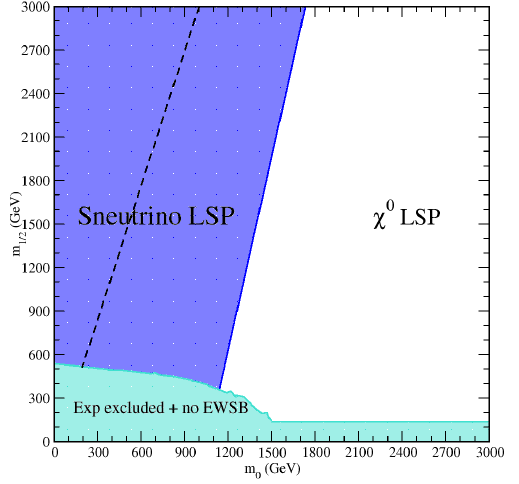}
\end{minipage}
\hspace{0.5 cm}
\begin{minipage}[t]{0.5\textwidth}
\vspace*{-5cm}
\caption{The $m_0$-$m_{1/2}$ plane for $\tan\beta=35$, $A_0=0$ and $\mu>0$. The dark blue area denotes the set of supersymmetric parameters where the sneutrino is the LSP in inverse-seesaw models. The white region has the neutralino as LSP in both standard and modified mSUGRA and the light blue region is excluded by experiments and theoretical constraints. On the left of the black dashed line the stau is the LSP in the standard mSUGRA.}
\label{fig:m0mh}
\end{minipage}
\end{figure}

In order to illustrate the mechanism we consider the simplest one-generation case, for simplicity. In this case where the sneutrino mass
matrix reads:
\begin{eqnarray}
 \mathcal{M}^2_{inv}  = 
\left (
\begin{array}{cc}
\mathcal{M}^2_+ & \mathbf{0}\\
 \mathbf{0} & \mathcal{M}^2_-
\end{array}
\right )
\end{eqnarray}
where the two sub--matrices $\mathcal{M_\pm}^2$ are:
\begin{eqnarray}\label{eq:snumatrix}
\mathcal{M_{\pm}}^2  = 
\left (
\begin{array}{ccc}
m^2_L+\frac{1}{2} m^2_Z \cos 2\beta+m^2_D & \pm (A_{\nu}v_2-\mu m_D {\rm cotg} \beta) & m_D M_R\\
 \pm (A_{\nu}v_2-\mu m_D {\rm cotg}\beta) & m^2_{N}+M_R^2+m^2_D & \mu_S M_R \pm B_{M_R}\\
m_D M_R & \mu_S M_R \pm B_{M_R} & m^2_S+\mu^2_S+M^2_R\pm B_{\mu_S}
\end{array}
\right )
\end{eqnarray}
in the CP eigenstates basis $\Phi^{\dag} = (\snu_{+}\,\tilde{N}_{+} \, \tilde{S}_+ \,\, \snu_- \,\tilde{N}_- \, \tilde{S}_-)$.
Once diagonalized, the lightest of the six mass eigenstates is our dark matter candidate and it is stable by $R$--parity conservation. Again the $\snu-Z$ coupling is reduced by the mixing with the right-handed field $\tilde{N}$ and is off-diagonal; moreover there is an additional fainting factor, due to the admixture with the sterile singlet $\tilde{S}$.

In the absence of the singlet neutrino superfields, the mSUGRA framework predicts the lightest supersymmetric particle to be either a stau or a neutralino, and only the latter case represents a viable dark matter candidate. In contrast, when the singlet neutrino superfields are added, a combination of sneutrinos emerges quite naturally as the LSP. A general analysis in the mSUGRA parameter space is shown in Fig. \ref{fig:m0mh}: the light blue area is excluded either by experimental bounds on supersymmetry and Higgs boson searches, or because it does not lead to electroweak symmetry breaking, while the region on the left of the black dashed line refers to stau LSP in the conventional mSUGRA case. As expected, in all of the remaining region of the plane (white region), the neutralino is the LSP as in the standard mSUGRA case. The new phenomenological possibility which opens up thanks to the presence of the singlet neutrino superfields where the sneutrino is the LSP corresponds to the blue area.

\begin{figure}[t!]
\vspace*{-1cm}
\begin{minipage}[t]{0.5\textwidth}
 \centering
\includegraphics[width=\columnwidth]{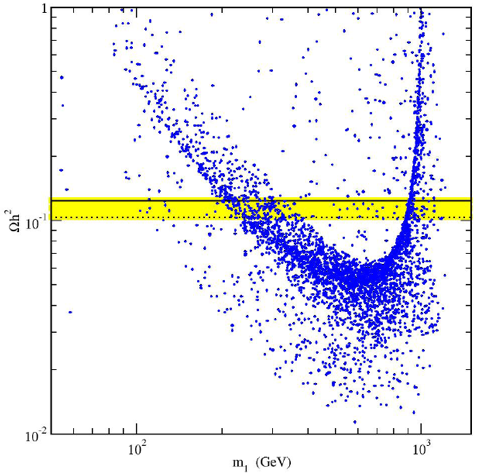}
\caption{Sneutrino relic abundance $\Omega h^{2}$ as a function of the sneutrino mass $m_{1}$ for a scan of the supersymmetric parameter space. The yellow band delimit the five years WMAP interval for CDM at $3 \sigma$ of C.L.: $0.104  \leq \Omega_{CDM} h^2 \leq 0.124$.}
\label{fig:om_inv}
\end{minipage}
\hspace{0.5 cm}
\begin{minipage}[t]{0.5\textwidth}
\centering
\includegraphics[width=\columnwidth]{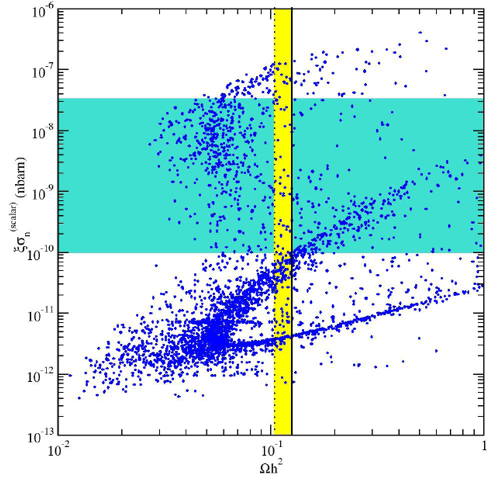}
\caption{ Sneutrino cross-section on nucleon $\xi\sigma_{nucleon}^{(scalar)}$ versus the relic density $\Omega h^2$ for a scan of the supersymmetric parameter space. The horizontal band denotes the current sensitivity of direct detection experiments; the vertical band delimits the $3 \sigma$ C.L. WMAP CDM range.}
\label{fig:sig_inv}
\end{minipage}
\end{figure}

The novelty of the spectrum implied by mSUGRA implemented with the
inverse seesaw mechanism is that it may lead naturally the lightest sneutrino $\tilde{\nu}_1$ as the LSP, instead of the
fermionic neutralino. The relic density of the sneutrino candidate is shown in Fig. \ref{fig:om_inv}: we see that a large fraction of the sneutrino configurations are compatible with the WMAP cold dark matter range~\cite{Komatsu:2008hk}, and therefore represent viable sneutrino dark matter models. In addition Fig. \ref{fig:sig_inv} shows that direct detection experiments do not exclude this possibility: instead, a large fraction of configurations are actually compatible and under exploration by current direct dark matter detection experiments. This fact is partly possible because of the inelasticity characteristics we have mentioned above, which reduces the direct detection cross-section to acceptable levels. The main feature of our non minimal extension of the MSSM is that the nature of the dark matter candidate, its mass and couplings all arise from the same sector responsible for the generation of neutrino masses. 

\section{Conclusions}
We have presented scenarios in which neutrino masses and dark matter arise from the same sector of the theory.
In the first section we have described the sneutrino phenomenology with a low scale Majorana mass in non minimal MSSM models with a standard seesaw for generating neutrino masses. The sneutrino turns out to be an interesting dark matter candidate, with the proper relic density and compatible with the direct searches of WIMPs. In the second section we show that an extended MSSM model within the inverse seesaw mechanism opens up new mSUGRA scenarios. Over large portions of the parameter space the model successfully accommodates light neutrino masses and sneutrinos dark matter with the correct relic abundance indicated by WMAP as well as direct detection rates consistent with current dark matter searches.

\section*{Acknowledgments}
This work was supported by IISN and Belgian Science Policy (IAP VI/11). 

\section*{References}
\bibliography{biblio}

\end{document}